# Study of Efficient Photonic Chromatic Dispersion Equalization Using MZI-Based Coherent Optical Matrix Multiplication

Sizhe Xing[(1)], Guoqiang Li[(1)], Ziwei Li[(1,2)], Nan Chi[(1,2)] and Junwen Zhang[(1,2)]*

[(1)] MoE Lab, Fudan University, Shanghai 200433, China, Junwenzhang@fudan.edu.cn
[(2)] Peng Cheng Lab, Shenzhen 518055, China

**Abstract** *We propose and study an efficient photonic CDE method using MZI-based coherent optical matrix multiplication. It improves the compensation performance by about 60% when the tap-length is limited, and only 50% taps of the theoretical value is needed for photonic CDE with 1-dB penalty.* ©2022 The Author(s)

**Introduction**

With the exponential growth of bandwidth demands, higher-speed information flow needed to be carried in the optical fibres. Coherent optics makes it possible for the high-speed, large capacity data transmission over long distances. In such optical coherent communication links, chromatic dispersion (CD) is one of the linear impairments that must be compensated [1]. Generally, the CD is compensated either by finite impulse response (FIR) filters in time-domain or inverse response in frequency-domain in digital signal processing (DSP). The number of FIR taps required is directly proportional to the transmitted distance, and has a quadratic relationship with the bandwidth. This results high energy consumption and high complexity by CD equalization (CDE) in the DSP over long-haul transmission. Digital CDE accounts for more than 20% of total DSP power consumption, and increases significantly with fibre distance [2].

Most recently, optical matrix-vector multiplications (MVM) are designed to address the growing demand for computing resources and capacity [3], among which the Mach–Zehnder interferometer (MZI)-based MVM is one of the most promising solutions [4]. Thanks to the coherent optics, MZI-MVM can represent positive and negative by phase, which is beneficial for subsequent design. Its ability in matrix multiplication has been proved more efficient and faster than electronic methods [5]. Linear transformations can be performed at the speed of light and detected at rates exceeding 100 GHz in MZI-assisted MVM and, in some cases, with minimal power consumption [6]. Therefore, it offers us a potential new method to address the CDE by photonic MVMs, thanks to the linear transformations in matrix multiplication.

In this paper, we propose and study the use of MZI-MVM for CDE, namely, photonic CDE, in coherent optical communication. To take advantage of MVM, the data is processed in blocks in our proposed scheme, which can transfer an $M \times M$ matrix as an $M^2 \times 1$ filter without reducing the operation speed or increasing the operation complexity. Besides, a novel taps reduction method is also proposed in this paper for the photonic CDE design. Simulation results confirm the effectiveness of photonic CDE, and the proposed filter design method can improve performance by 60% when the tap-length is limited. We have also verified that only 50% of the theoretical value of the number of taps needed to compensate for the dispersion of the same length. Finally, the photonic CDE by a 4×4 MZI-MVM is carried out for 200Gbps PDM-16QAM signals after fibre transmission and only about 1dB loss is observed compared to the digital CDE.

**Design of the photonic CDE**

CD is typically compensated for by a finite impulse response (FIR) butterfly structure, key part of which is the weight bank arranged in one column. However, the MZI-MVM is always designed as a square matrix, as this structure can address various computing scenarios. The weight bank of the MZI-MVM is arranged in few different columns. It is proved that these variform matrixes can perform the same convolution operation, ignoring the different arrangement. Here we use a 16taps FIR and a 4×4 matrix as an example in Fig.1 (a). As for the 16 taps filter, the input signal can be written as $X \in \mathbb{R}^{16 \times K}$, where $k$ represents the numbers of input signals, and each column follows the previous one by one symbol delays. To meet the input shape of the 4×4 MVM, $X$ can be separated into 4 blocks, $B_1, B_2, B_3, B_4$, in the direction of the columns. And then it can be rearranged as $X' = [B_1, B_2, B_3, B_4]^T$, where $B_i \in \mathbb{R}^{4 \times K}$ consists of 4 rows in $X$. Now the calculating process can be shown as:

$$diag(\sum_{i=1}^{4}\lambda_i X_i, \sum_{i=5}^{8}\lambda_i X_i, \sum_{i=9}^{12}\lambda_i X_i, \sum_{i=13}^{16}\lambda_i X_i)$$

$$= I \odot \begin{bmatrix} \lambda_1 & \lambda_2 & \lambda_3 & \lambda_4 \\ \lambda_5 & \lambda_6 & \lambda_7 & \lambda_8 \\ \lambda_9 & \lambda_{10} & \lambda_{11} & \lambda_{12} \\ \lambda_{13} & \lambda_{14} & \lambda_{15} & \lambda_{16} \end{bmatrix} \begin{bmatrix} X_1 & X_5 & X_9 & X_{13} \\ X_2 & X_6 & X_{10} & X_{14} \\ X_3 & X_7 & X_{11} & X_{15} \\ X_4 & X_8 & X_{12} & X_{16} \end{bmatrix} \quad (1)$$

The four output values at time i will be used to form the symbols $Y_i$, $Y_{i+4}$, $Y_{i+8}$, and $Y_{i+12}$. By simply adding the elements on the diagonal, we can get the result $Y = \sum_{i=1}^{16} \lambda_i X_i$ same as the output of a 16 taps FIR filter. It is worth noting that this process will not increase any complexity in the whole calculating process, the signal can be simply sent to the MVM one by one. And then, the result is calculated in real-time and output by a designed adder.

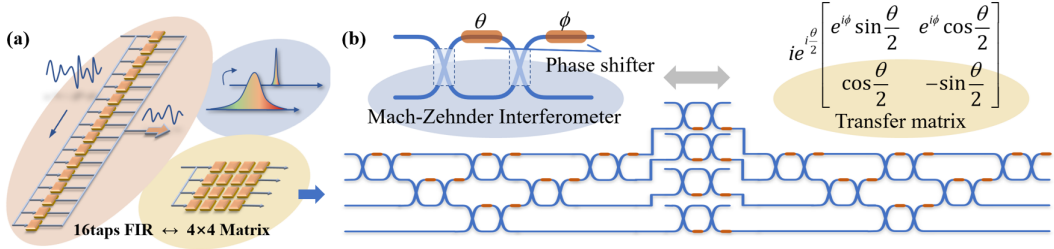

**Fig. 1:** (a) the transmission from FIR filter to a matrix; (b) the detailed constructure of a MZI-MVM.

Fig 1(b) shows the building block of the optical linear unit based on MZI and the constructed MZI-assisted matrix-vector multiplication. Each MZI consists of two 50/50 beam and two phase shifters. [7] proposes and demonstrates the MZI-based MVM method that each matrix can be firstly decomposed into two unitary matrices and a diagonal matrix and then each unitary matrix can be constructed by a series of MZI. Following this method, we construct a 4×4 matrix shown in Fig. 1 (b). and use it to compensate the CD.

**Photonic CDE simulation**
To evaluate the performance of the photonic CDE, we simulate the coherent optical fibre communication environment and use MVM for photonic CDE. In this process, the first problem is that MZI-MVM can only perform real number operations whereas CDE requires both inputs and filters to be complex. Therefore, we use the structure shown in Fig. 2 (a) to compensate this drawback. The input complex signal is firstly separated into Imaginary part and real part, and then sent to the MVM array which is composed of real and imaginary part computation. So the photonic CDE comprises four 4×4 MZI-MVMs. And then this module is used in the coherent fibre transmission system replacing the electronic CDE. The system structure diagram is shown in Fig. 2 (b). Here, we use two external cavity lasers (ECLs), both of which are with the same wavelength of 1551.9nm, to emit the signal light source and local oscilloscope (LO). After modulated by IQ-modulator, the 25G Baud signal is transmitted through various distance single mode fibre (SMF), and the signal is disturbed by various degrees of CD. In the receiver side, the signal and transmitted LO light are mixed and detected by an integrated coherent receiver (ICR) and undergoes a series of post-DSP. After resampled to 2 times the baud rate, then the signal of two samples for one symbol is transmitted in the photonic CDE. Here we also calculate the signal waveform that processed by the traditional electronic CDE and compare the waveform of these two methods. It is believed to be a good metric to measure the feasibility of photonic CDE. After that, the signal is than undergoes a series of equalizations and QAM demodulation. Finally, we get the resulting constellation and BER.

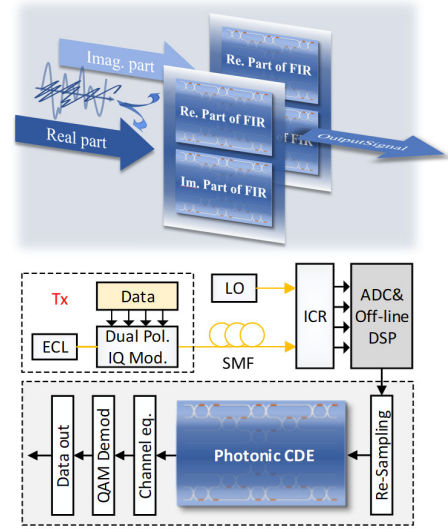

**Fig. 2:** (a) the photonic CDE constructure and (b) the setup of the simulation to test the performance of photonic CDE.

**Results and analysis**
Here we show how the filter is generated when the number of taps is fixed. Without limiting the number of taps, the precise needed number of taps can be calculated for the CD at different distances in Eq. (2):

$$h(n) = \sqrt{\frac{jcT^2}{D\lambda^2 z}} \exp(-j\frac{\pi cT^2}{D\lambda^2 z}n^2) \qquad (2)$$

Where $T$, $D$, $z$ and $\lambda$ respectively represent sample time, dispersion parameter, transmission distance and centre wavelength. In such a condition, CDE designing is a problem with unique solution. But the issue changes when the number of taps is fixed. We can no longer design filters with direct reference to formulas. In this paper, the method transforming from frequency domain filter to time domain convolution is adopted. When the number of theoretically required taps is less than the total number of elements in the matrix, the dispersion can be fully compensated. And with the transmission distance become longer than the range that the available MZI-MVM can compensate, the number of elements of the matrix is not enough, we can firstly generate a filter with a longer taps and then cut it down to the feasible range.

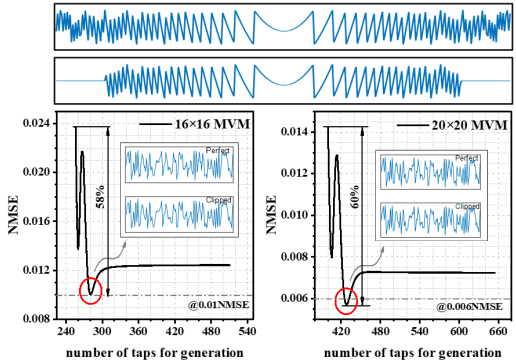

**Fig. 3:** the new method to generate the filter taps with fixed tap number

To verify the feasibility of this scheme, we carried out the experiment shown in Fig. 3 at the distance of 1000 and 1500km, respectively. The normalized mean square error (NMSE) between the waveform after a perfect filter and the waveform after the tap-number-fixed filter is used here as the standardize versus the initial generated taps. With the initial generated number of taps increases, the NMSE can quickly decrease about 60%. The waveform comparison is also shown in the insets. In these two cases as the transmission distances are up to 1000km and 1500km, the required tap number is 320 and 480, both compressed by about 20%.

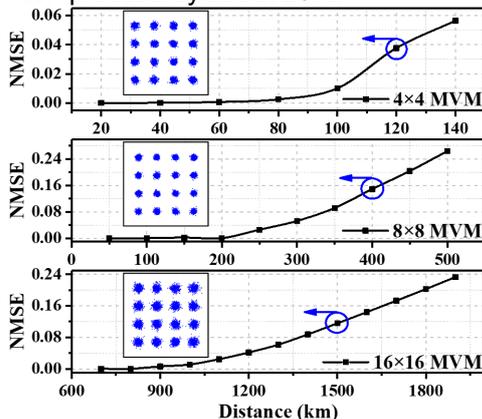

**Fig. 4:** NMSE of the waveform after tap-number-fixed CDE versus different distances.

We then numerically study the performance of NMSE versus different distances and the result is shown in Fig. 4. The insets reveal the final constellation. These three figures show the CD of various lengths that can be handled by matrices of various sizes, and we mark the limit distances. When distances increasing, the NMSE is initially around 0, and then it will grow slowly until the signal cannot be demodulated beyond a threshold. As for the 25G Baud signal, the maximum dispersion that each matrix can handle are 120, 400 and 1500 km, as the numbers of taps used are only the 42%, 50% and 53% of theoretical value. Referring to this result, numerical study confirms that the large-scale MZI-MVM can be used for photonic CDE of the distance up to thousands of kilometres.

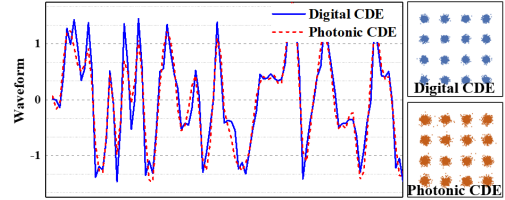

**Fig. 5:** the difference in the waveform lead by photonic CDE.

Finally, we construct a $4\times 4$ MZI-MVM and examine its performance for the 200Gbps PDM-16QAM signal. The signal through a photonic CDE will distorted by the phase noise and nonlinearity, which is totally different from the noise in coherent system and difficult to be equalized by the existed algorithm. The noise shows an irregular state, which also makes the calculation accuracy lower than the test case. Through the waveform in Fig. 5, we can find that the waveform after the photonic CDE is a little gentler than the digital CDE, which may lead by the response speed of the phase shifter. This result is tested as the fibre length of 80km, just in the work range of a $4\times 4$ MZI-MVM. And the two constellations respectively represent the received signal undergoes the digital CDE and photonic CDE. The estimated SNR is 18.5 dB for photonic CDE and 19.5 dB for digital CDE. A 1dB loss is caused by the noise of the MZI-MVM.

## Conclusion

We propose and study an efficient photonic CDE method using MZI-based MVM, transforming a $m\times m$ matrix into a $m^2\times 1$ filter. The method of designing CDE filter with tap number limited is proposed and demonstrated to be able to improve the compensation performance by about 60%. The photonic CDE by a 4×4 MZI-MVM for 200Gbps PDM-16QAM signals after fibre transmission achieves only about 1dB penalty compared with the digital CDE. Numerical study confirms that the large-scale MZI-MVM can be used for photonic CDE of the distance up to thousands of kilometres.

## Acknowledgements
This work is partially supported by National Key Research and Development Program of China (2021YFB2801804), National Natural Science Foundation of China (61925104, 62031011, 62171137), Natural Science Foundation of Shanghai (21ZR1408700), and the Major Key Project of PCL.